\documentclass[lettersize,journal]{IEEEtran}
\usepackage{amsmath,amsfonts}
\usepackage{algorithmic}
\usepackage{algorithm}
\usepackage{array}
\usepackage[caption=false,font=normalsize,labelfont=sf,textfont=sf]{subfig}
\usepackage{textcomp}
\usepackage{stfloats}
\usepackage{url}
\usepackage{verbatim}
\usepackage{graphicx}
\usepackage{cite}
\usepackage{paralist}   
\usepackage{wrapfig}
\usepackage{float}
\usepackage{tabularx}
\usepackage{makecell}   
\usepackage{amssymb}
\usepackage{xcolor}
\usepackage{multicol}
\usepackage{multirow}
\usepackage{booktabs}
\usepackage{hyperref}

\begin{document}

\title{ChartGPT: Leveraging LLMs to Generate \\ Charts from Abstract Natural Language}

\author{Yuan Tian,
        Weiwei Cui,
        Dazhen Deng,
        Xinjing Yi,
        Yurun Yang,
        Haidong Zhang,
        Yingcai Wu

\thanks{
Y. Tian, D. Deng, X. Yi, Y. Yang, and Y. Wu are with Zhejiang University. 
E-mail: \{yuantian, dengdazhen, yixinjing, yurunyang, ycwu\}@zju.edu.cn. 
Dazhen Deng and Yingcai Wu are the corresponding authors.
}
\thanks{W. Cui and H. Zhang are with Microsoft. E-mail: \{weiwei.cui, haidong.zhang\}@microsoft.com.}
\thanks{Manuscript received xxx, xx, 2024; revised xxx, xx, 2024.}}

\markboth{Journal of \LaTeX\ Class Files,~Vol.~14, No.~8, August~2021}%
{Shell \MakeLowercase{\textit{et al.}}: A Sample Article Using IEEEtran.cls for IEEE Journals}


\maketitle

\begin{abstract}
The use of natural language interfaces (NLIs) to create charts is becoming increasingly popular due to the intuitiveness of natural language interactions. 
One key challenge in this approach is to accurately capture user intents and transform them to proper chart specifications. 
This obstructs the wide use of NLI in chart generation, as users' natural language inputs are generally abstract (i.e., ambiguous or under-specified), without a clear specification of visual encodings.
Recently, pre-trained large language models (LLMs) have exhibited superior performance in understanding and generating natural language, demonstrating great potential for downstream tasks.
Inspired by this major trend, we propose ChartGPT, generating charts from abstract natural language inputs. 
However, LLMs are struggling to address complex logic problems. 
To enable the model to accurately specify the complex parameters and perform operations in chart generation, we decompose the generation process into a step-by-step reasoning pipeline, so that the model only needs to reason a single and specific sub-task during each run. 
Moreover, LLMs are pre-trained on general datasets, which might be biased for the task of chart generation. 
To provide adequate visualization knowledge, we create a dataset consisting of abstract utterances and charts and improve model performance through fine-tuning. 
We further design an interactive interface for ChartGPT that allows users to check and modify the intermediate outputs of each step. 
The effectiveness of the proposed system is evaluated through quantitative evaluations and a user study.
\end{abstract}

\begin{IEEEkeywords}
Natural language interfaces, large language models, data visualization.
\end{IEEEkeywords}

\section{Introduction}
\label{sec:introduction}
Natural language interfaces (NLIs) have become a popular interactive strategy for data analysis and visualization creation~\cite{narechania2020nl4dv, luo2021ncNet}.
For example, a user can easily create a histogram showing the distribution of IMDB ratings for a movie dataset by simply saying ``create a histogram showing the distribution of IMDB ratings.'' 
Compared to traditional methods, NLIs provide a shortcut for analysts not proficient in visualization programming, such as D3.js or Vega-Lite~\cite{2017-vega-lite}, to create visualizations.
Even for senior visualization users, NLIs can free them from tedious programming issues or interactive editings on visualization toolkits (e.g., Tableau~\cite{tableau}).

The key of NLIs is to precisely capture user intents and generate appropriate visualizations under the ambiguity and underspecification of natural languages.
While experts in visual analytics are capable of specifying all necessary information for visualization generation in one utterance, including data fields, data transformations, chart types, and visual encodings, beginners in visualization programming may struggle to provide all the information.
Demonstrated by previous studies~\cite{rose2004understanding, setlur2019arklang}, user queries are underspecified in many cases.
For instance, the utterance ``What type of movies make the most money?'' implicitly refers to the field of ``gross profit.'' 
The term ``type'' can be understood differently (e.g., genre, rating, etc.) in various contexts. 
Such ambiguity makes it hard to map utterances to concrete chart specifications. 
Traditional methods combine lexical parsing and predefined rules to support abstract inference to some extent~\cite{setlur2019arklang}.
For example, NL4DV~\cite{narechania2020nl4dv} facilitates attribute inference from computing the similarity with data fields, values, and defined aliases and enables task and chart type inference through predefined rules. 
However, such methods are limited by the ability of parsers to understand natural language. 
In addition, the predefined aliases and rules can be hard to maintain, modify, and expand~\cite{wu2021ai4vis}.

Recently, large language models (LLMs), such as Bert~\cite{devlin2018bert}, GPT-3~\cite{brown2020GPT3} and ChatGPT~\cite{chatgpt}, have demonstrated outstanding performance in natural language understanding.
These models, pre-trained on a massive corpus of text, have acquired a vast amount of knowledge and can be utilized for various downstream tasks~\cite{wei2022emergent}, such as data transformation~\cite{huang2023nl2rigel}, narration generation~\cite{chung2022talebrush, ying2023livecharts}, and web design~\cite{kim2022stylette}. 
The remarkable success of these LLM applications inspires us to investigate their potential for visualization generation. 
However, using LLMs to generate visualizations from abstract utterances presents two main challenges.

\textbf{Controlling chart parameters with LLMs.} 
The process of visualization generation involves complex parameters and operations. 
Users have to specify parameters such as mark, field, encoding, and aggregation, which are then rendered by visualization systems (e.g., Vega-Lite and Tableau) to transform the original data table and produce the chart. 
While language models (LLMs) can generate fluent and informative answers to human questions, they may not always be accurate, which is well-known as the  ``hallucination problem''~\cite{welleck2019neural_hallucination}. 
This makes it challenging to use LLMs directly in visualization generation, as a single incorrect parameter could negatively impact the subsequent operations and potentially compromise the entire process.
To tackle this challenge, we adopt a systematic approach by breaking down the chart generation process into a series of interrelated sub-tasks, following the principle of least-to-most idea~\cite{zhou2022least}.
This decomposition allows us to leverage the strengths of LLMs to produce well-defined and manageable outputs for complex parameters and operations involved in chart creation.

\textbf{Lacking approaches to inject visualization knowledge.} 
LLMs are designed and trained to handle general language-related tasks, such as text generation, recognition, and summarization.
To make LLMs more domain-specific, two methods are commonly used: prompting and fine-tuning.
Prompting refers to providing the model with a text that includes the context of domain tasks and expected outputs.
Although effective, this approach is not always practical, especially when the model needs to be provided with a large amount of knowledge (e.g., Draco rules in our scenario) in a single prompt.
Fine-tuning the LLM with appropriate datasets can provide more examples and knowledge. 
While there are well-established datasets in NL2VIS~\cite{srinivasan2021collecting, luo2021synthesizing}, these datasets mainly consist of explicit natural language descriptions or cover limited datasets, which are not suitable for our scenario.
To address this challenge, we constructed a dataset of abstract utterances with corresponding charts.
The dataset enables LLMs to learn user intents in visual data analysis and generate chart configurations with the desired formats.

In this study, we introduce ChartGPT, leveraging LLMs to generate charts from abstract utterances.
We broke down the chart generation process into a series of sub-tasks for the LLM to solve sequentially and constructed an abstract utterance dataset to fine-tune the model (FLAN-T5-XL~\cite{Hyung2022flant5}).
Based on the fine-tuned model, we developed an interactive interface that allows users to explore and modify the intermediate steps of chart generation. 
We evaluated our proposed method through quantitative experiments and a comparative user study with the state-of-the-art NL2VIS methods. 
We also summarized the feedback from the usability study and discussed future work for improving the system.
The main contributions of this study are as follows.
\begin{itemize}
    \item We propose a framework to generate charts from abstract utterances using fine-tuned LLMs. 
    \item We construct a dataset of abstract utterances and charts for LLM fine-tuning. The dataset could facilitate future machine learning research in this direction.
    \item We conduct quantitative experiments and user studies to prove the usefulness of the proposed method. The feedback could shed light on future applications of LLMs in visualizations.
\end{itemize}
\begin{figure*}[htbp] 
  \centering
  \includegraphics[width=\linewidth]{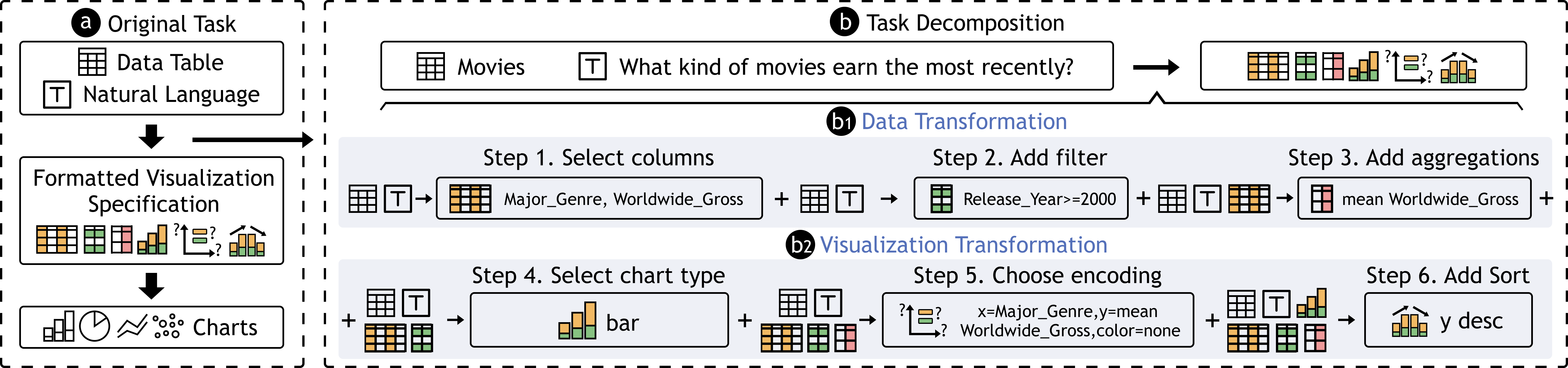}
  \caption{
An example of chart generation problem formulation. 
(a) The task comprises three stages: input context (data table and natural language), formatted visualization specification, and charts. 
(b) We decompose the first stage transformation process into two successive transformations: data transformation (b1) and visualization transformation (b2), involving six steps. 
At each step, the model utilizes the input context and previous answers to generate the next output.
  }
  \label{Fig:reasoning_example}
\end{figure*}

\section{Related Work}

\subsection{Visualization Recommendation}

Recently, there has been a growing interest in exploring visualization recommendation techniques that can assist data workers in tackling the laborious task of creating visualizations~\cite{qin2020making, DBLP:journals/tvcg/LinCWL23, ying2023metaglyph, wei2022evolutional, zhou2023intelligent}.
These techniques are mainly classified into two categories: rule-based and machine learning (ML)-based~\cite{saket2018beyond, wu2021ai4vis}. 
Rule-based methods map data to visual encoding according to visualization knowledge, such as the conclusions from empirical studies. 
A large number of recommendation systems, such as APT~\cite{mackinlay1986APT}, Show Me~\cite{mackinlay2007showMe}, CompassQL~\cite{wongsuphasawat2016compassQL}, and Voyager~\cite{wongsuphasawat2015voyager1, wongsuphasawat2017voyager2}, are compiled from visualization rules.
To improve the usability of visualization rules, Moritz et al.~\cite{moritz2018Draco} translated the rules into answer set programming and formulated a knowledge base.
Though effective, rule-based methods might suffer from flexibility, as the manually specified rules by experts are difficult to update, modify, and maintain. 
This limits their adaptability to diverse data types or changing conditions.

In contrast, ML-based methods have the advantage of being able to learn from data and adapt to changing conditions, making them more flexible and robust~\cite{dong2023VItcevis, sun2022VIlearning, jiang2022VIvisualizations}. 
For example, DeepEye~\cite{luo2018deepeye} and Draco-learn~\cite{moritz2018Draco} use machine learning algorithms to rank recommended visualizations based on visualization design rules. Other studies, such as Data2Vis~\cite{dibia2019data2vis} and Table2Charts~\cite{zhou2021table2charts} utilizes sequence-to-sequence models to map datasets to visual representations. 
KG4Vis~\cite{li2021kg4vis, zhu2021VIvisualizing} uses knowledge graphs to support explainability for recommendations.
To generate multiple-view visualizations~\cite{SONI2024dashboard}, MultiVision~\cite{wu2022multivision} and Dashbot~\cite{deng23dashbot} adopt deep learning methods to model datasets.
These studies primarily focus on creating visualizations from data tables. 
In this work, we take a more challenging approach, comprehending natural language intentions to generate charts.

\subsection{Natural Language Interfaces for Data Visualization}

Natural language interfaces are proven efficient in specifying data visualizations~\cite{shen2021nli, voigt2022nli, chen2021nebula}. 
Many studies utilized semantic or lexical parsing techniques to infer user intent and generate visualizations. 
Articulate~\cite{sun2010articulate} extracted visual tasks and attributes and selected visualizations with a graph reasoner algorithm.  
DataTone~\cite{gao2015datatone} proposed interactive ambiguity widgets to help users resolve ambiguity in natural language. 
FlowSense~\cite{yu2019flowsense} utilized semantic parsers to assist dataflow diagram construction. 
Users can expand and adjust dataflow diagrams via natural languages. 
Eviza~\cite{setlur2016eviza} employed a probabilistic grammar-based approach and allowed an interactive query dialog with an existing visualization. 
Evizeon~\cite{hoque2017Evizeon} further applied language pragmatics principles to support visual analytical conversations. 
NL4DV~\cite{narechania2020nl4dv} incorporated lexical and dependency parsing techniques to infer attributes and tasks from user utterances and generated visualizations. 
With recent advancements in natural language processing, attempts have been made to utilize deep learning-based language models to produce visualizations.
For example, ncNet~\cite{luo2021ncNet} employed a Transformer-based sequence-to-sequence model to convert natural language queries into visualizations.

However, these studies mainly aim at explicit requests and are difficult to deal with incomplete or implicit utterances. 
Some studies, such as NL4DV, enable implicit attribute inference by computing the similarity with data fields, values, and defined aliases. 
Ask Data~\cite{setlur2019arklang} resolved partial utterances based on syntactic and semantic constraints and produced an intermediate language to generate visualizations. 
However, The performance of these methods is greatly limited by the capability of the language parsers.
Additionally, the predefined aliases and rules may hinder flexibility, as they are hard to maintain, modify, and expand~\cite{wu2021ai4vis}.
In this paper, we aim to utilize the language comprehension ability of pre-trained LLMs to tackle the challenge of abstract utterances.

\subsection{Large Language Models for Data Analysis}
Recently, there have been significant advancements in large language models (LLMs), such as online models GPT-3~\cite{brown2020GPT3} and GPT-4~\cite{openai2023gpt4}, as well as open-source models flan-T5~\cite{Hyung2022flant5} and LLaMa~\cite{touvron2023llama}. 
Pre-training on tens of TB of text data, LLMs have demonstrated superior performance in understanding and generating natural language. 
These models have been applied to various domains, including data transformation~\cite{huang2023nl2rigel}, narration generation~\cite{chung2022talebrush, ying2023livecharts}, and web design~\cite{kim2022stylette}.

Specifically, recent studies have explored utilizing LLMs for data analysis. 
Some studies employ LLMs to generate visualization code (e.g., Python and Vega-Lite) directly. 
For example, CHAT2VIS~\cite{maddigan2023chat2vis} generates visualization code in Python by prompting LLMs with table schema, column types, and utterances. 
Similarly, LIDA~\cite{dibia2023lida} defines visualization generation as a four-stage generation problem and leverages GPT-3.5 to generate visualization code. 
Other studies explored a broader application of LLMs in data analysis. 
GPT4-Analyst~\cite{cheng2023isgpt} proposes a framework that utilizes prompts to direct GPT-4 in performing data collection, visualization, and analysis. 
Data-Copilot~\cite{zhang2023datacopilot} can generate requests, select the needed interfaces, and invoke the corresponding interface tools sequentially or in parallel. 
All of these works are based on prompt engineering and depend on online models such as GPT-3.5 and GPT-4, which are not fully controllable and stable~\cite{welleck2019neural_hallucination, openai2023gpt4}. 
These models might suffer from inherent hallucination problems that occasionally provide unstable output with incorrect answers, leading to failure to follow the designed pipeline.

Different from the above methods that use generic LLMs, we opt to train a visualization-specific LLM to address the problem of chart recommendation.
Specifically, we adopt the chain-of-thought~\cite{wei2022chain, zhou2022least} idea to decompose the task and then solve it sequentially. 
Instead of relying on prompt engineering, we fine-tuned an open-source LLM on our constructed abstract utterances dataset. 
Additionally, we developed a template for the model input and output, enhancing parsing and applicability across various visualization representations, with Vega-Lite serving as an example. 
\begin{figure*}[htbp] 
  \centering
  \includegraphics[width=\linewidth]{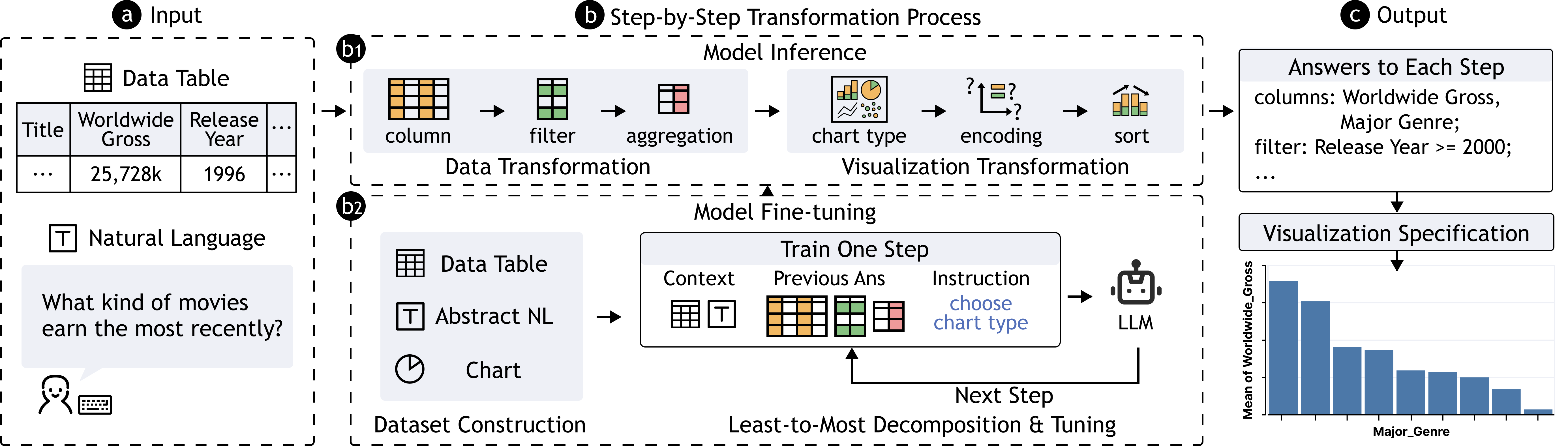}
  \caption{
  ChartGPT overview. 
  ChartGPT takes a data table and an utterance provided by the user as input (a). 
  To generate the chart, ChartGPT employs a step-by-step transformation process (b) that decomposes the chart generation task into six sequential steps (b1). 
  Each step is solved by the LLM fine-tuned on our constructed dataset (b2).
  By leveraging the output from each step, ChartGPT generates visualization specifications and presents charts to the user (c).
}
  \label{fig:teaser}
\end{figure*}

\section{Background and Problem Formulation}
\label{section:Problem Formulation}

This section introduces the background of reasoning strategies and describes how we formulate the chart generation problem into step-by-step reasoning sub-tasks. 

\subsection{Reasoning Strategies in LMs}
\label{subsection:reasoning}

For language models (LMs), reasoning is defined as the process of breaking down a complex task into simpler sub-tasks for LMs to handle effectively~\cite{mialon2023augmented}. 
Specifically, in the least-to-most reasoning strategy~\cite{zhou2022least}, the original task is divided into a sequence of sub-tasks, starting with the simplest and gradually increasing in complexity. 
Through the reasoning process, LMs can solve more complicated sub-tasks with the help of previously solved sub-tasks.

We also adopt a decomposition approach to tackle the chart generation task. 
We formulate the task as a fixed sequence of sub-tasks and tackle them with an LLM that generates an answer based on the problem context and the outputs from previous sub-tasks. 
Notably, all sub-tasks are handled by the same model, as LLMs possess the capability to generalize across various tasks.
Finally, the answers from all sub-tasks are consolidated to produce a complete chart. 

\subsection{Problem Formulation}

We formulate our problem based on the Information Visualization Data State Reference Model~\cite{chi2000taxonomy,card1999readings}, which outlines the visualization pipeline as a sequence of data stages and explains how data undergoes various transformations from one stage to the next. 

As illustrated in~\autoref{Fig:reasoning_example}a, we formulate the problem into three data stages: table data, formatted visualization specification, and charts. 
Specifically, a formatted visualization specification is a text sequence that satisfies a specific visualization grammar and can be parsed and rendered into a chart.
Examples include Vega-Lite~\cite{2017-vega-lite}, Vega-Zero~\cite{luo2021ncNet}, and the chart templates defined in Table2Charts~\cite{zhou2021table2charts}. 
We also propose a formatted template compatible with our method and pipeline. 

Our key challenge is the first transformation of stages: generating visualization specifications from table data based on user utterances. 
We decompose the process into a series of sub-tasks and formulate each sub-task as a formatted sequence-to-sequence problem, illustrated in~\autoref{Fig:reasoning_example}b. 

\subsubsection{Problem Decomposition}

Inspired by grammars of graphics~\cite{wilkinson2012grammar, harris1999information, mackinlay1987automatic}, we divided the process from data to visualization specification into two successive transformations: data transformation and visualization transformation (\autoref{Fig:reasoning_example}b1 and \autoref{Fig:reasoning_example}b2). 
Both consist of three sub-tasks, resulting in a sequence of six sub-tasks performed step-by-step. 

\textbf{Data transformation. } Data transformation contains operations that deal with table data. 
After this transformation, the transformed data can be encoded directly into visual channels. 
The data transformation process includes three sub-tasks: selecting columns, filtering rows, and adding aggregations. 
\textbf{First}, relevant columns are selected based on the data and user utterance, usually involving 1-3 columns. 
\textbf{Second}, data rows are filtered if needed. 
\textbf{Third}, data columns are aggregated using functions such as \textit{count}, \textit{average}, and \textit{sum} if necessary.

\textbf{Visualization transformation. } After obtaining the transformed data, we should determine the appropriate encoding of visual channels. 
This process also contains three sub-tasks: choosing chart type, determining visual encoding, and adding optional operations. 
\textbf{First}, the model needs to infer which chart type is suitable for the selected data, aggregation, and utterance. 
\textbf{Second}, the model is required to map the data fields to visual channels. 
Notably, the fields in this sub-task have been transformed. 
For example, if executing counting on a specific field ``a'', the field to be encoded should be ``count(a)''.
\textbf{Third}, there are possibly optional operations for the resulting charts, such as color, sorting order, and bin width, etc.
In this study, we primarily consider sorting by axis for simplicity. 

After the six-step successive transformation process, adequate information is obtained to formulate a visualization specification. 
Indeed, chart generation extends far beyond the aforementioned steps. 
Other design alternatives encompass factors such as color, size, bandwidth, and orientation.
The transformation involved in chart generation is not limited to simple filter conditions and aggregation functions as well. 
These alternatives can be realized through engineering extensions, i.e., introducing additional steps or options and expanding related datasets. 
In this study, as a proof-of-concept system, we only consider the six sub-tasks and several main design choices to simplify the problem and encourage more design alternatives to be explored in future work.

\subsubsection{Answer Template for Sub-tasks}

After decomposing the problem into sub-tasks, the next step is to model each sub-task as a sequence-to-sequence problem solvable by the LLM.

\begin{figure}[htbp] 
  \centering 
  \includegraphics[width=\linewidth]{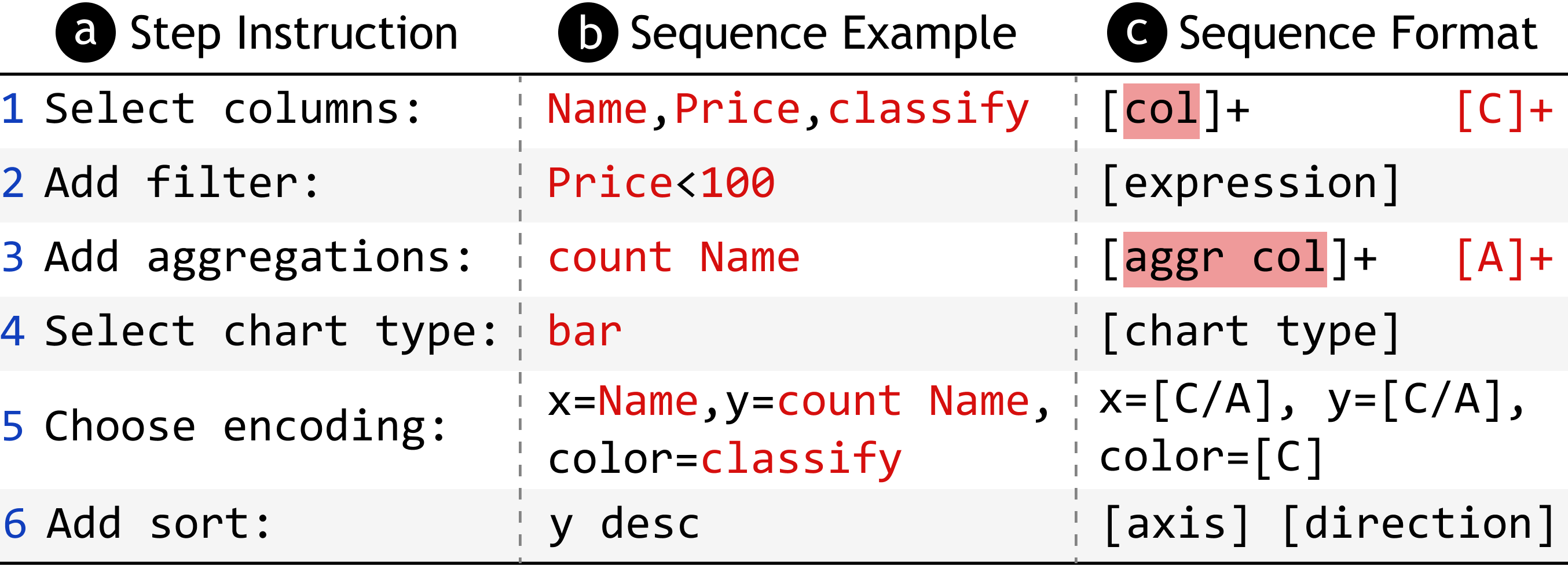}
  \caption{
  The template sequence for each sub-task. 
  }
  \label{Fig:template} 
\end{figure}

As illustrated in~\autoref{fig:teaser}, the model processes each sub-task with a text sequence input, consisting of the utterance, table data, and answers to previous sub-tasks to enhance reasoning. 
The model will output a text sequence as the sub-task answer, expected to meet two criteria:
(1) cover all mandatory information and (2) be well-formatted to enable accurate parsing and valid specification construction. 
To support this, we defined a corresponding template sequence for each sub-task, as illustrated in~\autoref{Fig:template}, similar to Vega-Zero~\cite{luo2021ncNet}. 

Specifically, the \textbf{selected columns} are represented by the data fields (denoted by `col') separated by commas. 
\textbf{Filter} is an expression string comprised of conditions, each involving a data field with a predicate such as equal, greater than, and less than. 
These conditions can be logically connected by `and' / `or.' 
\textbf{Aggregation} functions (denoted by `aggr') can be applied to the selected columns as `aggr col', including count, average, sum, max, and min. 
\textbf{Mark} specifies the chart type, including bar, pie, line, and scatter. 
\textbf{Encoding} maps the axes with selected columns (C) and aggregations (A). 
\textbf{Sort} indicates which axis (x/y) to sort and in which order (desc/asc). 

The model outputs the answers to each sub-task and combines the answers of filter, mark, encoding, and sort to generate a Vega-Lite specification. 
Some sub-tasks, including filter, aggregation, color encoding, and sort, may not always be necessary, and the model will output `none' in these cases. 

Our system covers seven chart types commonly found in data analysis~\cite{battle2018beagle,vartak2015seedb}: bar, stacked bar, line, grouped line, scatter, grouped scatter, and pie. 
More complex chart types, such as radar charts and heat maps, are not included in the template. 
Filter and aggregation also contain design options that are beyond our scope. 
In our work, we only focus on the basic design alternatives for each sub-task to initially validate the potential of LLMs to reason about visualization design. 

\section{ChartGPT}

This section describes the approach utilized to guide the LLM's reasoning for answering each sub-task.
We derived an abstract NL2VIS dataset to fine-tune a large language model and generate the answers through the model. 
The dataset was constructed through prompting GPT-3~\cite{brown2020GPT3}. 
The model~\footnote{https://huggingface.co/yuan-tian/chartgpt} is released on Hugging Face. 
The dataset, prompts, and model input settings are provided in our supplementary materials.

\subsection{Model Input}

For a specific sub-task, since its answer is based on the input context and the answers of previous sub-tasks, the model input should comprise three pieces of information: (1) table data, (2) user utterance, and (3) answers to previous sub-tasks. 
However, due to the limited token length that LLMs can handle, it is not feasible to feed the entire table data into the model. 
Thus, we only incorporate the column names and the top two data rows into the model input.
Moreover, to compensate for the possible model cognitive bias from including partial data only, we added the type of each data column to the input to provide a data overview.

\subsection{Reasoning Prompt and Abstract Utterances}
\label{subsection:reasoning_abstract}

An effective way to use an LLM for a specific downstream task is to design a prompt that guides the model in understanding the task target. 
The prompt can comprise both instructions and examples. 
For instance, when a task is to classify the sentiment of Tweets, the prompt may include an instruction that states ``decide whether a Tweet's sentiment is positive, neutral, or negative,'' along with a few examples such as ``I loved the new Batman movie! =\textgreater positive''.
The model should then be able to generate a response of ``negative'' for ``I hate chocolate.''
This technique of including examples in the prompt is called few-shot prompting~\cite{brown2020GPT3}. 
Few-shot prompting can facilitate the model to understand the context and task, which motivated us to consider whether this technique can be applied to generate visualizations from utterances.

However, due to the flexibility of natural language, the user utterances can be abstract for different information and on different levels. 
For example, in terms of \textbf{information abstraction}, users may omit the chart type or refer to the data fields in vague terms, such as using ``popular'' to represent the column ``rating'' or ``gross''. 
For \textbf{level abstraction}, users may concretely express their visualization intent, such as ``A pie chart showing the number of faculty members for each rank.'', which directly specifies the selected columns (rank), aggregations (count) and chart type (pie chart). 
On the other hand, they may also use more abstract queries, such as simply saying ``show rank.''
This omission of specifications can lead to multiple interpretations and reasoning paths for a particular sub-task. 
For instance, the choice of chart type can be determined by the selected columns (e.g., a scatter plot for two quantitative attributes) or the analytical intent of the user (e.g., a histogram for phrases like ``distribution''). 

The complexity of interpretation and reasoning paths makes it challenging to provide sufficient examples for each sub-task within a single prompt. 
To assist the model in gaining a more comprehensive understanding of the sub-task interpretations, we construct a dataset and fine-tune the model accordingly.

\subsection{Dataset for Fine-tuning}
\label{subsection:dataset}

\subsubsection{Dataset Requirements}

The dataset to fine-tune our model should consist of (data, utterance, chart) triplets. 
To provide the model with sufficient knowledge, the dataset should cover diverse interpretation and reasoning paths. 
Therefore, the dataset should meet several requirements: 

\textbf{Various domains and types of data and charts. }
Ensure diverse data sources across various domains to avoid overfitting to a single domain. 
If the domains are too concentrated, for example, if most tables are related to movies, the model may overfit this context, making it less adaptable to data from other domains. 
In addition, the data types and chart types involved should also be comprehensive and diverse. 

\textbf{Different levels of information for data analysis.} The utterances should be abstract for different information and on different levels, as is mentioned in \autoref{subsection:reasoning_abstract}. 
It should also cover various expressions, such as the way to describe selected columns (e.g., explicit or implicit) and phrasing (e.g., command, question, or query). 

Previous work about NL2VIS datasets includes Quda~\cite{fu2020quda}, NLV Corpus~\cite{srinivasan2021collecting} and nvBench~\cite{luo2021synthesizing}. 
Quda consists of 14,035 user utterance queries covering various analytical tasks. 
However, no associated charts are provided. 
NLV Corpus collected 893 utterances involving ten chart types and further analyzed the utterance features, spanning different expressions and abstractions. 
However, NLV Corpus is based on only three data tables, making it overly concentrated. 
The nvBench dataset is the closest to matching our requirements, with 25,750 (data, utterance, chart) triplets from 105 domains of table data. 
However, most utterances in nvBench are very explicit~\cite{cheng2023isgpt}.
Therefore, we construct our dataset based on nvBench, which consists of utterances in different abstractions and expressions.

\subsubsection{Dataset Construction}

To construct a dataset based on nvBench, the main task is maintaining the diverse data tables and visual design and generating abstract utterances from the original triplets. 
To maintain the diversity, we randomly select part of the original triplets covering all domains and chart types, etc.
To generate abstract utterances, we use GPT-3 (text-davinci-003) and involve four co-authors to verify their correctness. 
We produce the dataset in the following process:

\textbf{Charts selection. }We select charts from nvBench to align with our requirements. 
First, as nvBench contains some charts involving multiple tables (using the `join' operation), we remove this part of the data.
Second, nvBench consists of (data, utterance, chart) triplets from various domains and chart types, categorized into four hardness levels: easy, medium, hard, and extra hard. 
These hardness levels reflect the complexity of chart generation. 
For instance, a chart that encodes three columns and requires filter, aggregation, and sort operations may be classified as extra hard.
We select the charts randomly and ensure the selected data covers all domains, hardness levels, and chart types with a relatively balanced proportion.

\textbf{Abstract utterance generation. }
After selecting the charts, we use GPT-3 to generate abstract utterances for each chart from its corresponding (data, utterance, chart) triplets. 
For each triplet, we manually design a prompt to guide GPT-3 to do this: 
First, we provide the top few lines of the CSV table data and describe a scenario in which we develop a tool to generate charts automatically based on user utterances and table data. 
Then, we give an original utterance from the triplet as an explicit utterance example. 
We tell the GPT-3 model that we need abstract utterances to test the tool's performance, and require the model to generate abstract utterances based on the explicit original utterance and the table data. 
We also guide the model that the generated utterances should be more natural, vague, and incomplete and can be in various phrasings. 

Moreover, we dynamically checked the diversity of generated utterances during the generation process. 
For example, at first, we observed that the results tended to use many polite and verbal expressions, such as ``Can you show me'' (e.g., ``Can you show me the amount of matches for each competition on a graph?'') and ``I want to see'' (e.g., ``I want to see a visualization of the number of cinemas in different locations, please.''). 
This may be attributed to GPT-3's interpretation of ``natural'' as incorporating polite and verbal expressions. 
While these phrasings are commonly used, NLV Corpus demonstrates that short queries or commands are also very often in users' utterances. 
Examples from NLV Corpus include ``histogram for creative type'' and ``Plot IMDB rating against Rotten Tomatoes rating.'' 
As NLV Corpus classified the majority of utterances into query, question, and command, we modified the prompts to accommodate a range of phrasings and obtained utterances without overly polite and verbal expressions such as ``Budget creation trend'' and ``Plot capacity by opening year''.
We retained the previously generated utterances and included them alongside the new additions in our final dataset. 

\textbf{Abstract utterance correction.} 
The generated utterances should remain consistent with the original chart in the (data, utterance, chart) triplet from nvBench. 
In other words, the chart should be a reasonable answer to the utterance. 
As most generated abstract utterances remove or blur some information from the original utterance, some of them became inconsistent with the original charts. 
Specifically, for filters, compared to chart types and other settings that may still stay consistent with the original chart even after being removed, utterances that remove filter information are no longer consistent with the original chart. 
Generally, the inconsistent data were filtered manually through three co-authors before being reviewed by another co-author. 
Any disagreements in the correction of the data were resolved through discussion.

\textbf{Step-by-step answer generation. } As our model outputs consist of the answers to the intermediate sub-tasks, we need to parse the chart configuration and extract the answers to each sub-task. 
We then combine the answers and the formatted template to construct the expected output of the model. 

\subsubsection{Dataset Statistics}

\begin{figure}[htbp] 
  \centering
  \includegraphics[width=\linewidth]{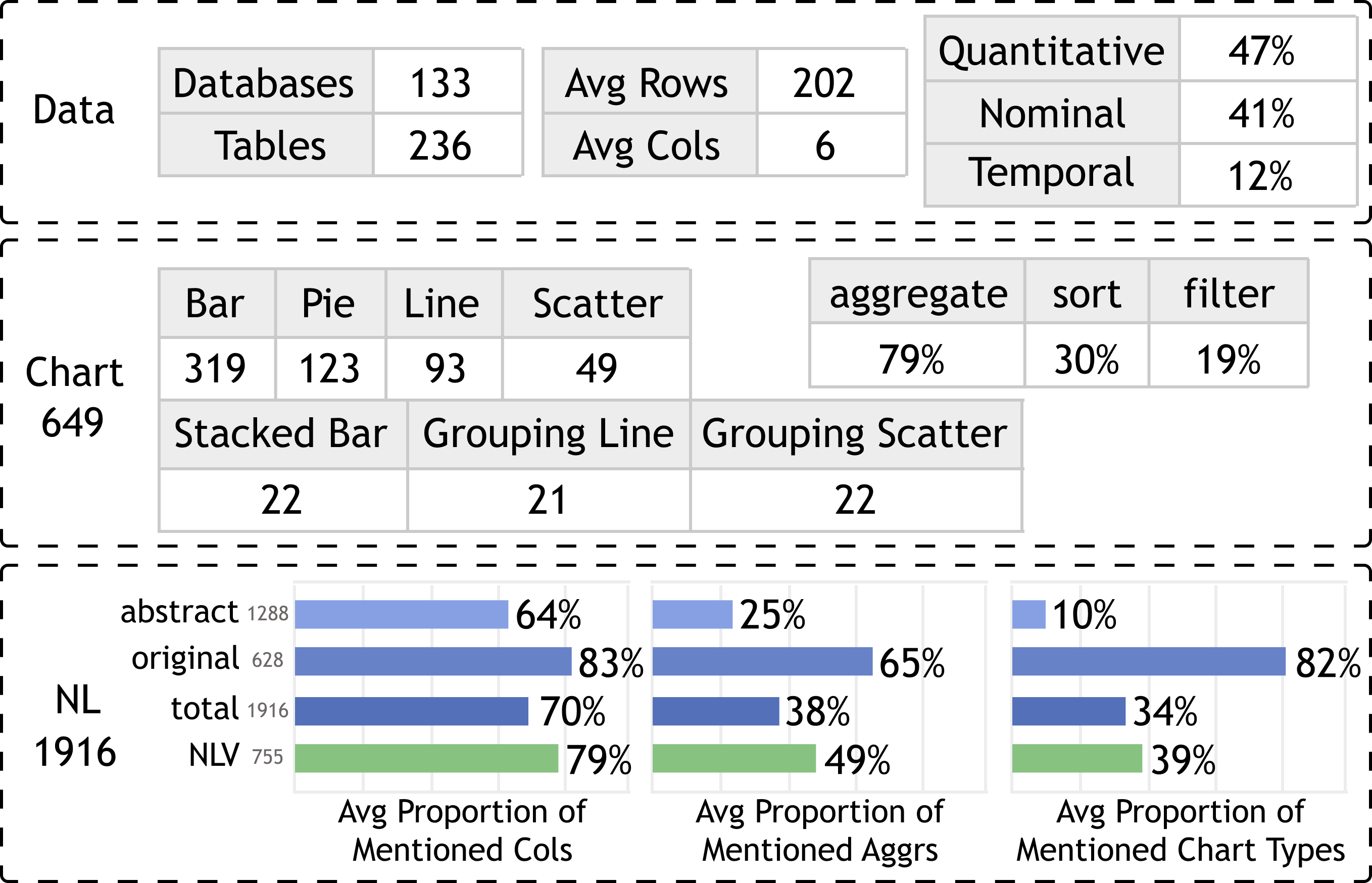}
  \caption{The statistics of our constructed dataset. Specifically, ``abstract'' denotes our generated abstract utterances, ``original'' denotes the maintained original utterances from nvBench, and ``total'' denotes our total dataset, which includes the ``abstract'' and ``original'' utterances.}
  \label{Fig:statistics}
\end{figure}

Our constructed dataset consists of 1,916 (data, chart, utterance) triplets, including 236 data tables, 649 charts, and 1,916 utterances. 
\autoref{Fig:statistics} illustrates the statistics of our dataset. 

\textbf{For data tables}, our dataset contains 236 tables from 133 databases, with an average of 5 columns and 202 rows.
Quantitative columns account for 47\%, nominal columns 41\%, and temporal columns 12\%.
\textbf{For charts}, our dataset covers seven chart types. 
Specifically, 79\% of charts involve aggregation, 30\% involve sorting, and 19\% involve filtering operations.

\textbf{For utterances}, we retained the original utterances from nvBench. 
The final dataset comprises 1,916 utterances, with 1288 newly generated abstract utterances and 628 original ones.
Furthermore, we compared the statistics between our dataset and the human-created dataset, NLV Corpus~\cite{srinivasan2021collecting}. 
We quantified the frequency of explicit information related to selected columns, aggregations, and chart types mentioned in the utterances. 
\textbf{For selected columns}, we calculated the proportion of explicitly mentioned column names. For example, if a chart involved three columns, but the corresponding utterance only referred to two of them, the proportion would be 2/3.
\textbf{For chart types and aggregations}, we examined the presence of explicit expressions, such as ``bar'', ``scatterplot'' for chart types, and ``number of'', ``count'' for aggregations.

The results indicate that among NLV Corpus utterances, selected columns are explicitly mentioned more frequently (79\%), whereas chart types (49\%) and aggregations (39\%) are often omitted or vaguely expressed. 
The utterances from nvBench have a higher occurrence of explicit information across the board, particularly for aggregations (65\%) and chart types (82\%).
However, after fusing with the abstract utterances generated with GPT-3, the resultant dataset exhibits a significant reduction in explicit information, particularly concerning aggregations and chart types, which closely resemble NLV Corpus. 
As a result, the constructed dataset looks natural and similar to the human-created ones to some extent.

\begin{figure}[htbp] 
  \centering
  \includegraphics[width=\linewidth]{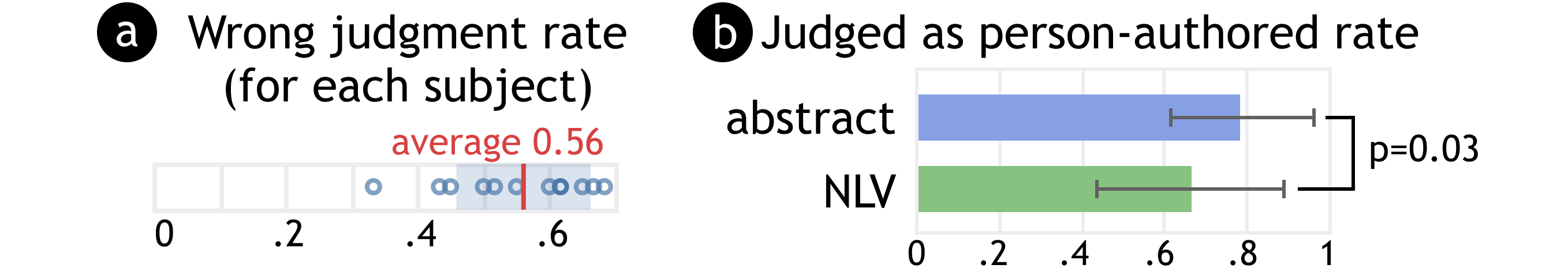}
  \caption{The Turing test results between our generated utterances and NLV Corpus ones. (a) The rate of wrong judgment of each subject. (b) The average rate of the two sets that were judged as human-created.}
  \label{Fig:turing}
\end{figure}

In the analysis above, we focus on assessing the explicit mentions of columns, aggregations, and chart types, as they can be quantified with less ambiguity. 
However, quantifying the explicit mentions of encoding, filtering, and sorting can involve the subjective opinions of different people.
To further measure the quality of our generated abstract utterances and assess their proximity to human-created utterances, we conducted a Turing test. 

\subsubsection{Turing Test}

We recruited 14 subjects (7 males and 7 females, all of whom possessed experience in data analysis) to conduct a Turing test evaluating the quality of our generated utterances. 
We randomly selected 30 utterances from NLV Corpus across 3 tables and 30 utterances from our generated abstract utterances involving 8 tables with shuffled order. 
During the test, each subject was provided with an utterance alongside the corresponding table at a time. 
The scenario presented to the subjects was as follows: ``Imagine a tool that automatically generates charts based on the table and users' utterances. Which of the utterances below might be created by a real user?''
We explicitly informed the subjects that some displayed utterances were human-created and some were not. 
Their task was to distinguish between the two categories based on two perspectives: (1) the naturalness of the phrasing and (2) the meaningfulness of the context.
We hypothesized that the rate of the generated abstract utterances judged as human-created would be at the same level as the NLV Corpus. 
After the experiment, we compensated each subject with \$5. 

\textbf{Overall results.} 
The results revealed an average error rate of 56\% (\autoref{Fig:turing}a), with the lowest error rate recorded at 33\%, suggesting that it was hard for subjects to distinguish between the GPT-generated utterances and human-created ones. 
Additionally, we computed the average rate ($\alpha$) at which each utterance was judged as human-created.
The overall average $\alpha$ for all 60 samples was 0.73, indicating that subjects labeled most samples as human-created.

\textbf{Comparison between generated utterances and human-created ones. }
Comparing the two sets, the average $\alpha$ values for our generated abstract utterances and NLV Corpus ones were 0.79 and 0.67, respectively (\autoref{Fig:turing}b). 
The corresponding standard deviation (SD) values were 0.17 and 0.23. 

To evaluate the disparity, we conducted a Mann-Whitney U test, which indicated a significant difference (p = 0.03 \textless 0.05). 
This result suggested that the generated utterances were even more likely to be perceived as human-created than the NLV Corpus ones.
To understand this discrepancy, we examined the NLV Corpus samples with lower $\alpha$ values.
One utterance stood out with a significantly low $\alpha$ of 0.14: ``Sum(Sales) by Order Date split by Category render line asc''. 
This utterance is similar to captions in format, which is considered less natural by most subjects.
NLV Corpus also acknowledged that their collected utterances contain such samples whose phrasing was relatively infrequent.

\subsection{Model Fine-tuning}
\label{subsection:fine-tuning}

\begin{figure*}[!htb] 
  \centering
  \includegraphics[width=\linewidth]{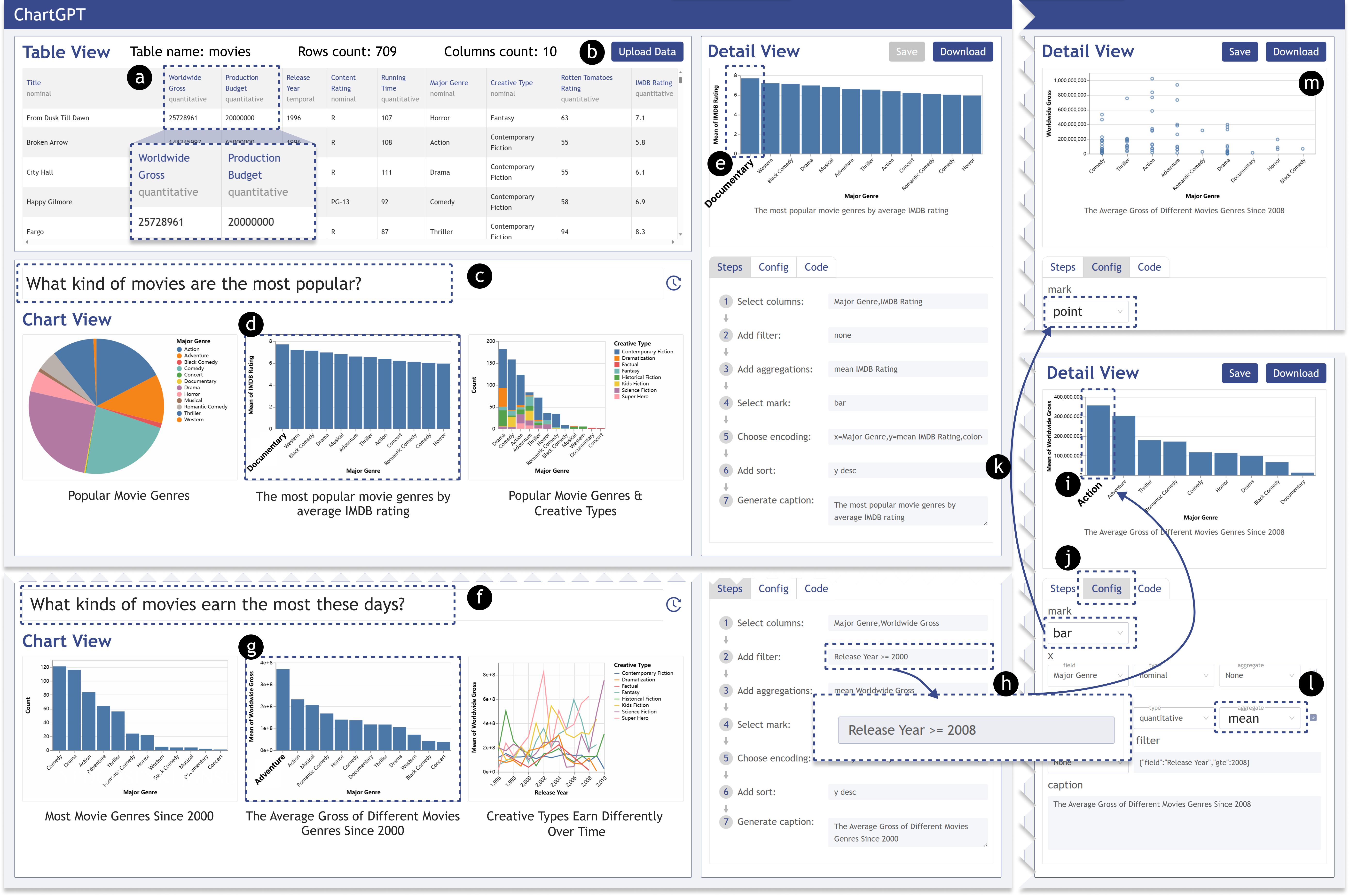}
  \caption{
  ChartGPT interface consists of three views: table view (a-b), chart view (c-g), and detail view (h-m). 
  Table view displays the data table and relevant data features. 
  Chart view enables users to input their utterances and presents the generated charts. 
  Detail view provides chart specifications and allows users to modify the results through interactions. 
  }
  \label{Fig:interface}
\end{figure*}

We first divided our dataset into a training set consisting of 1,538 triplets for fine-tuning and a test set with 378 triplets (invisible to the model) for evaluation (4:1 split).
Then, we fine-tuned the open-source FLAN-T5-XL model~\cite{Hyung2022flant5} with the AdamW optimizer~\cite{loshchilov2019adamw} on the training set. 

We selected Flan-T5 as it has undergone pre-training on various tasks and possesses strong reasoning capabilities.
We employed a learning rate of 1e-4, a global batch size of 16, and trained for five epochs.
Generally, the trained model obtains an evaluation loss of 0.10. 
These parameters are chosen based on the model document~\footnote{https://huggingface.co/docs/transformers/model\_doc/t5}, trial and error, and the capacity of our computational resources.
We show the evaluation results in \autoref{section:evaluation}. 

\subsection{Top-k Charts Generation}

ChartGPT is designed to generate a set of top-k charts (k=3 by default) in response to a given utterance. 
We incorporate two strategies to produce the top-k valid charts efficiently.

\textbf{To remove invalid candidates}, 
for the candidates generated by the model, we identify and eliminate the invalid candidates that contain
(1) column names not present in the data, 
(2) filter expressions that are grammatically wrong or can not be applied to the data,
and (3) aggregation functions, chart types, encoding channels, and sort tokens falling outside our valid space. 
\textbf{To generate results efficiently}, we adopt the beam search~\cite{sutskever2014sequence} to retain the top combinations of the candidates based on cumulative probabilities.
Finally, we return the top-k candidates to generate the charts. 
\section{Interface}

We developed an interface with three views: table view, chart view, and detail view. 
We present the features of our interface through a usage scenario based on a movie dataset.

To begin, the user uploads the CSV file (\autoref{Fig:interface}b). 
The data table is displayed with the column types, including nominal, quantitative, or temporal. 
The user then quickly navigates through the columns, types, and relevant data. 
S/he notices that the table contains 10 columns and 709 rows, each row providing information about a particular movie. 

The user wants to know ``what kind of movies are the most popular?'' and enters this question into the search box (\autoref{Fig:interface}c). 
ChartGPT then returns the top three charts based on the input. 
The user observes that the first and the third charts display the number of movies by genre and creative types, respectively, and the second chart shows the average IMDB rating of each genre. 
The user is interested in the second one (\autoref{Fig:interface}d) and understands that the movie genre with the highest average IMDB rating is Documentary. 

In addition to the count and ratings, the user further notes that the data contains information on gross and budget (\autoref{Fig:interface}a).
The user changes the input to ``What kinds of movies earn the most these days?'' (\autoref{Fig:interface}f). 
The results update, and the second and the third charts are about worldwide gross. 
The user investigates the charts and checks them in the detail view. 
S/he observes that the second chart (\autoref{Fig:interface}g) has a filter condition of ``Release Year \textgreater= 2000'', which corresponds to the utterance ``these days''.

The user is not fully satisfied with the filtering condition and expects more recent movies. 
S/he changes the condition to ``Release Year \textgreater= 2008'' and regenerates the result from step 3 (\autoref{Fig:interface}h). 
After re-rendering, the user discovers that the genre with the highest average gross since 2008 is Action (\autoref{Fig:interface}i). 
Furthermore, the user wants to see the distribution of movies for each genre. 
Therefore, the user switches to the config mode (\autoref{Fig:interface}j) in the detail view. 
S/he changes the mark type to ``point'' (\autoref{Fig:interface}k) and removes the aggregation of the y-axis (\autoref{Fig:interface}l), resulting in a scatter plot that meets the needs (\autoref{Fig:interface}m).
\section{Evaluation}
\label{section:evaluation}

\begin{figure}[!htb] 
  \centering
  \includegraphics[width=\linewidth]{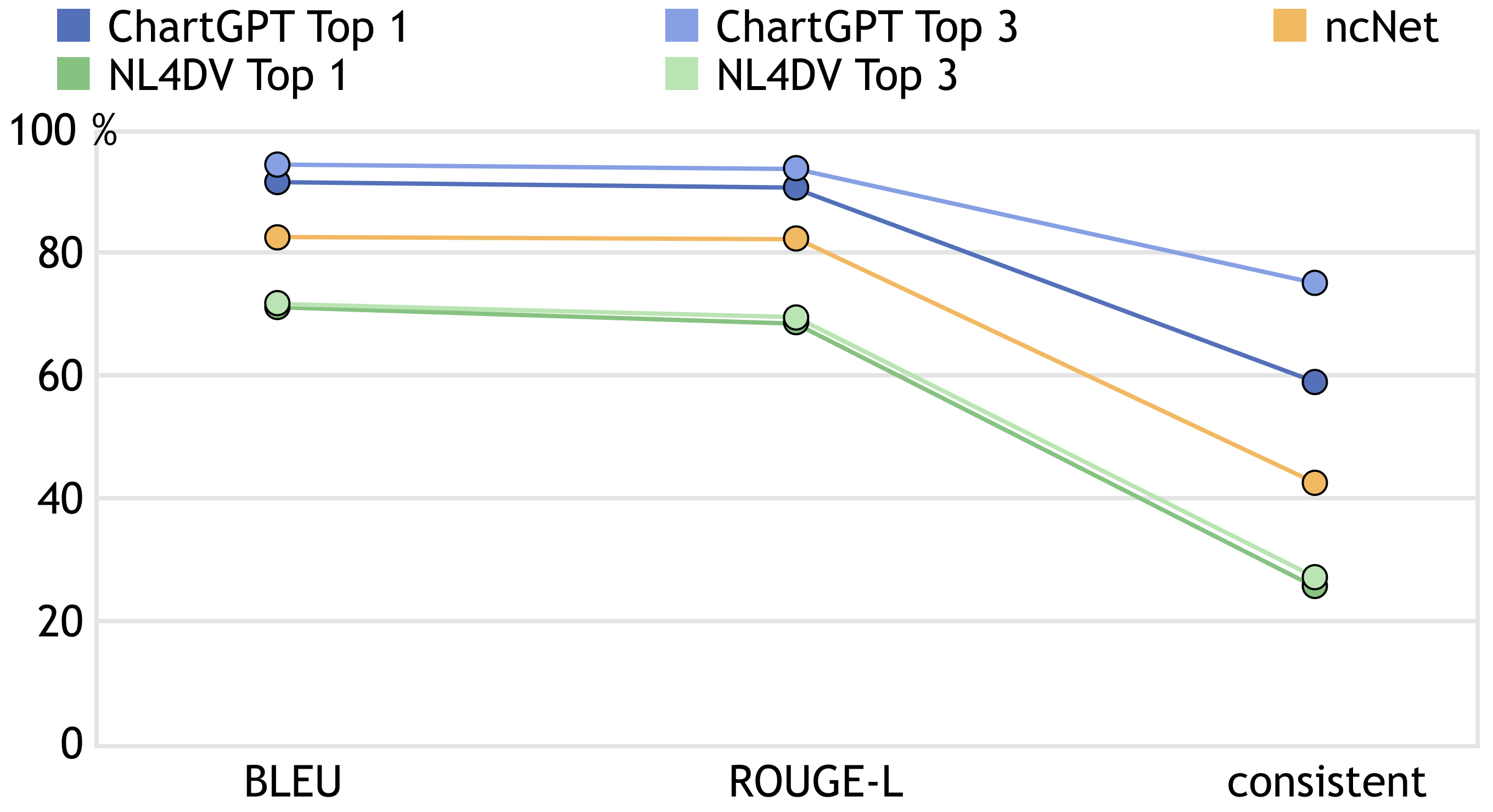}
  \caption{The evaluation result shows the performance of ChartGPT, ncNet, and NL4DV on different metrics. }
  \label{Fig:evaluation}
\end{figure}

This section introduces the quantitative evaluation of ChartGPT with NL4DV and ncNet.

\subsection{Evaluation Setup}

We used our test set (different from the training set) derived in~\autoref{subsection:fine-tuning} to evaluate the performance of ChartGPT, NL4DV, and ncNet. 
As both our system and NL4DV can return more than one result, we reported the top-1 and top-3 results for these two methods and reported the top-1 result for ncNet. 
However, please note that the design spaces of the three methods are also slightly different from each other. 
For example, NL4DV supports boxplots and tick charts but doesn't support pie charts. 
For fairness, we only compared results that can be produced by all methods. 
For the test data with configurations in our design space that NL4DV does not support, we didn't introduce them into the result statistics. 

\subsection{Evaluation Metrics}

We measured two metrics: consistency and similarity. 
Consistency is used to count how many results are exactly the same as the ground truth.
In addition, as abstract utterances may cause ambiguity, we further accounted for how similar the results are to the ground truth. 
We hypothesized that even if the utterance is ambiguous and can correspond to multiple correct answers, these answers are also similar to ground truth to some extent. 
Therefore, we used the degree of similarity to further measure the system's ability to handle abstract utterances. 

\textbf{Consistency Metrics.} We define a result as ``consistent'' if the result is identical to the ground truth. 
In our scenario, ``identical'' means identical in all supported design alternatives, including mark, encoding, aggregation, sort, and filter. 
In addition, we consider two scatter plots with x and y reversed as consistent as well, as they still point to equivalent results~\cite{srinivasan2021collecting}.

\textbf{Similarity Metrics.} We define the ``similarity'' of a result as the degree to which it is similar to the ground truth in terms of the design alternatives. 
We converted the ground truth and results of different methods into equal-length word sequences, and then compared the similarity of the sequences. 
The format of the sequence is defined as an 8-word sequence, i.e., [mark] [x field] [x aggregation] [y field] [y aggregation] [color field] [filter] [sort], and each part is a single word. 
Then, we measured the ROUGH-L~\cite{lin2004rouge} and BLEU~\cite{papineni2002bleu} metrics between the results and the ground truth sequences. 
ROUGH-L calculates the similarity based on the length of the longest common subsequence (LCS), which is affected by both the value and order of words.
Under this metric, if the selected fields and aggregations in both the ground truth and the model result are the same but encoded on different axes, the score will reduce. 
We suppose that charts with the same selected fields and aggregations but mapped to different axes from the ground truth are still acceptable when compared to the ones with inappropriate encoding. 
Therefore, as a complement to ROUGE-L, we measured the BLEU score, which allows the model to switch the order of some encoded fields. 

Specifically, before converting into a sequence, we validate whether the result is related to the data table and can be parsed properly into a Vega-Lite format. 
For example, if the result contains a column name that is not in the table, it is considered invalid. 
Parsing such results into Vega-Lite and displaying them will report errors or undefined displays because it cannot find the corresponding data. 
We marked the similarity and consistency of such results as zero. 

\subsection{Evaluation Results}

Our evaluation results are presented in \autoref{Fig:evaluation}, which showcases the top-1 and top-3 reviews of ChartGPT, ncNet, and NL4DV. 
The results indicate that ChartGPT outperforms the other two approaches in terms of both the consistency metric and similarity metric, with its top-1 and top-3 reviews scoring higher than those of ncNet and NL4DV. 

\textbf{Comparison with the baselines. }
Looking through the tested cases, there are two key factors that account for the differences between the approaches: 
One is semantic understanding. 
ChartGPT has a better parsing of the semantic information of the columns and utterances.
Examples include inferring the column ``sex'' from ``male and female'', column ``age'' from ``how old'', and inferring a temporal column and count aggregation from ``when create the most departments''. 
The other is omitted information. 
Abstract utterances often omit information such as aggregation and chart types, which requires the reasoning capability of the visualization specifications. 
ChartGPT is based on Flan-T5, which is previously fine-tuned on chain-of-thought (CoT) reasoning tasks and is further fine-tuned by us on visualization datasets in a CoT way, so it may have a better reasoning ability of omitted information. 

\textbf{
Metric difference analysis. 
}
The consistency metric is drastically lower when compared to the other two metrics, which is possibly due to two factors. 
First, ambiguity in abstract utterances often results in multiple reasonable answers. 
For instance, consider the abstract utterance ``How many documents are at each location? '' from the original utterance ``Show the number of documents for each location code in a pie chart''. 
This abstraction removes the chart-type information, making a bar chart also a reasonable response.
Second, partial correct inferences occur when the model misses some subtle yet critical to chart expressiveness information.
For example, the model may correctly extract the needed columns but give wrong aggregations, or miss the filter and sort conditions.

\section{User Study}

We derived a comparative study and a usability study to evaluate ChartGPT further. 
Through the user studies, we want to 
(1) compare the results of ChartGPT with the two baseline methods from users' perspective, and (2) evaluate the usability of ChartGPT. 

\subsection{Comparative Study}

In this study, we recruited 12 subjects (6 males and 6 females, all of whom possessed experience in generating data visualization) to conduct a comparative study evaluating the quality of generated charts from different approaches (ChartGPT, ncNet, and NL4DV). 
None of them has the experience of using the approaches above. 

\textbf{Tasks and Data. } 
We sampled 15 utterances from the test set, corresponding to 13 data tables and 42 charts generated from NL4DV (top-1), ncNet, and ChartGPT (top-1).
Subjects were presented with the tables, utterances, and generated charts in random order and were required to compare and rank the quality of charts, deciding which charts were more reasonable for the table and utterances based on their preferences. 
If a chart makes no sense in their opinion, it won't be included in the ranking. 
The sampling is based on two steps. 
First, we selected the tables that are close to common sense to ensure that subjects can understand the context. 
Second, we selected the abstract utterances from the selected tables and ensured that (1) the utterances are in various abstractions and phrasings and (2) the chart types are all included. 

\textbf{Procedure. } The entire experiment lasted about 10-25 minutes. 
First, subjects were introduced to the background of NLIs for generating data visualizations for 3 minutes. 
Then, they began to compare and rank the quality of generated charts based on the provided data table and utterances. 
Subjects were required to ensure that they understood the data table and utterances before performing the actions. 
They were allowed to ask about the meaning of the data table, utterance, or a particular legend in the chart but had to rank the charts entirely according to their own preferences. 
After the experiment, we compensated each subject with \$5.

\textbf{Results. } 
We counted the ranking results of the subjects. Specifically, for the user's ranking of the charts corresponding to a particular utterance, we normalized the rankings into scores from 0 to 3, with the first ranking scored as 3 and the charts that did not appear in the ranking scored as 0. 
Additionally, we calculated the proportion that each approach was first-ranked. 
We used a Friedman Test to examine whether a significant difference exists across the approaches and a post hoc Wilcoxon Test to compare the pair-groups. 

\begin{figure}[htbp] 
  \centering 
  \includegraphics[width=\linewidth]{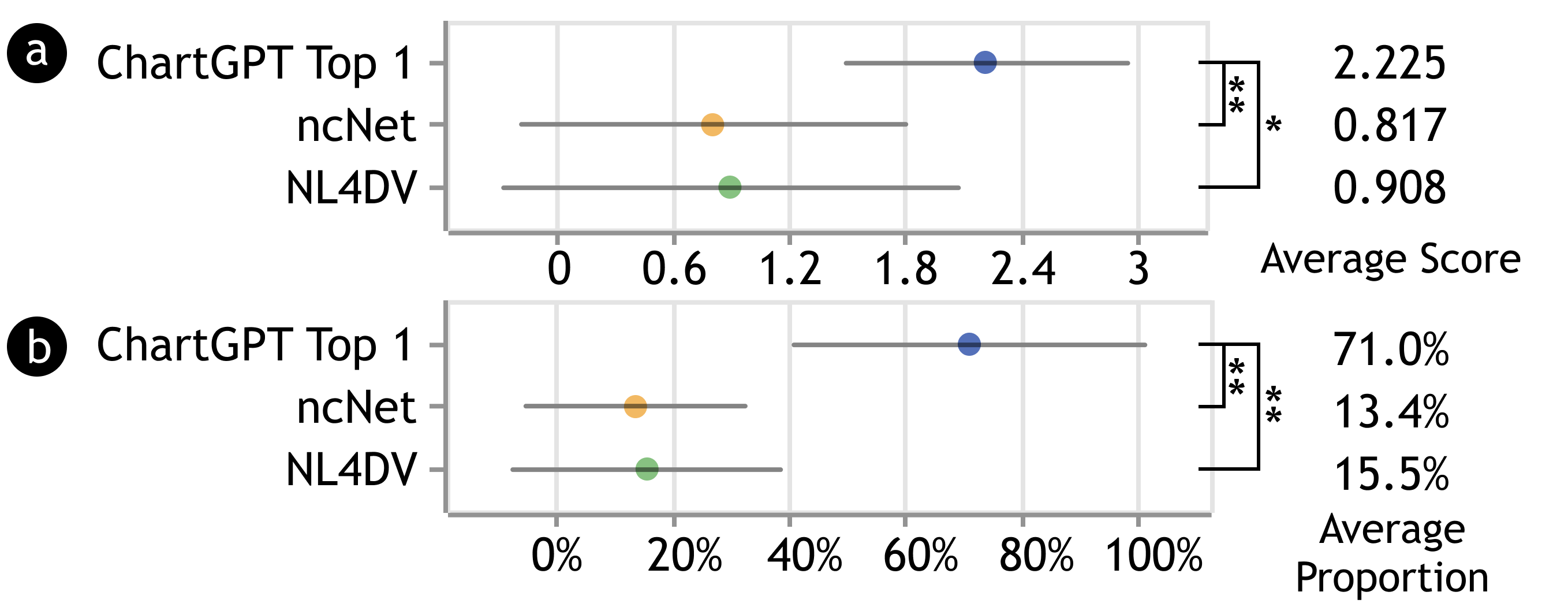}
  \caption{Results of the comparative study with SD values (*: p $<$ 0.05, **: p $<$ 0.01), including (a) average ranking scores and (b) first-ranked proportions.}
  \label{Fig:stage1} 
\end{figure}

The results (\autoref{Fig:stage1}) showed both significant differences in the ranking score ($\chi^2$=8.00, p $<$ 0.05) and first-ranked proportion ($\chi^2$=17.64, p $<$ 0.001). 
Overall, ChartGPT had the best performance (i.e., the higher ranking score and first-ranked proportion on average, p $<$ 0.05). 

\subsection{Usability Study}

\begin{figure*}[!htb] 
  \centering 
  \includegraphics[width=\linewidth]{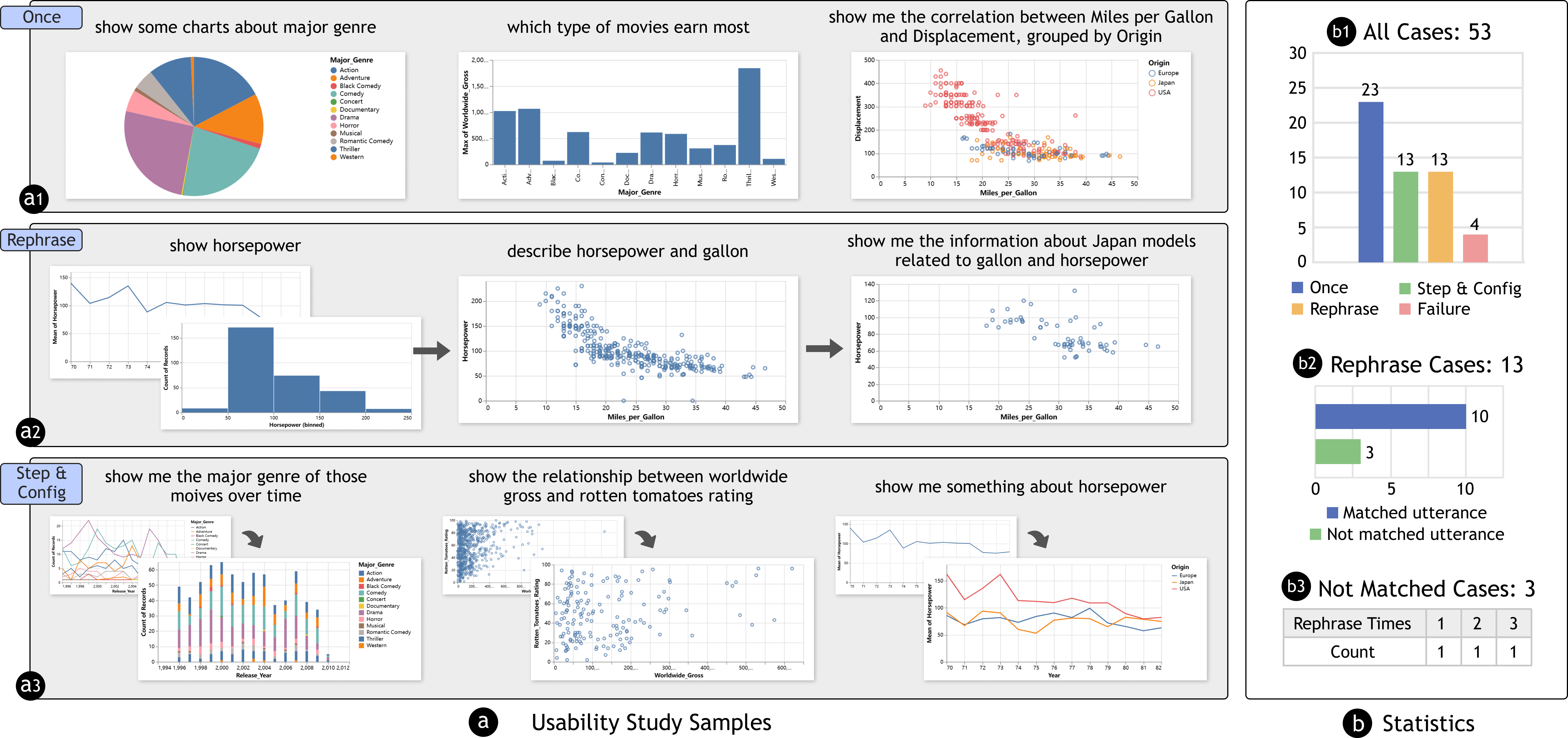}
  \caption{
  Results of the usability study, including samples of generated charts from subjects (a) and quantitative statistics (b).
  }
  \label{Fig:stage2} 
\end{figure*}

\subsubsection{Experiment Settings}

\textbf{Participants. } We recruited 12 subjects (S1-S12, 6 males and 6 females) from different departments, including Computer Science (3), Sports Science (2), Digital Media Design (2), Urban Informatics (1), Industrial Design (1), Geographic Information Science (1), Agricultural Engineering (1) and Corporate Finance (1). 
Most subjects were familiar with data visualization, with an average self-reported score of 3.4 on a 5-point Likert Scale. 
All subjects had experience using tools to author charts, including Microsoft Excel, Vega-Lite, D3.js, G2, ECharts, and Matplotlib. 
In addition, all of them had experience in using natural language interfaces (including ChatGPT) and scientific English writing. 

\textbf{Tasks and Data. }
Subjects were provided with two data tables (movies and cars) and were required to choose the one they were more familiar with or more interested in. 
They were required to explore the selected data with ChartGPT and create at least four desired charts. 
During the creation process, if the default generated chart did not match their desires, subjects could rephrase their input, modify the step answers and regenerate the results, or modify the chart configuration directly. 
However, if subjects could not get the desired chart no matter what action they took, or if the desired chart was not supported, they could give up the intent and try to generate another one. 
In total, the created charts should contain at least two chart types and involve at least three different data columns. 
Both the movies and cars data come from NLV Corpus~\cite{srinivasan2021collecting} and have more than 9 columns and 300 rows, involving all three types of values (temporal, nominal, and quantitative). 
We chose these two data tables as their context is close to common sense and is easy to understand. 

\textbf{Procedure. }
The entire experiment lasted about 20-35 minutes. 
Subjects were first presented with the movie and car datasets and were required to select one of them based on familiarity or interest.
We ensured that subjects could understand the dataset before the next process.
Subjects were then introduced to the interface and interactions of ChartGPT.
During the introduction, we did not provide the subjects with any concrete input examples to avoid biasing their language organization. 
Instead, we encouraged users to formulate their own input and introduced the interface and interactions along the way.
After the introduction, subjects began to create their desired charts with their selected data.
All inputs and actions taken by subjects are recorded.
Finally, we interviewed the subjects to collect their feedback about ChartGPT. 
After the experiment, we compensated each subject with \$10.

\subsubsection{Quantitative Results}

The results of the usability study are illustrated in \autoref{Fig:stage2}, and the corresponding statistics are presented in \autoref{Fig:stage2}b. 
A total of 53 historical logs were collected from the subjects, and 49 of them resulted in successfully generated charts. 
The other 4 failed logs indicated that the subjects could not obtain a satisfactory chart, thus giving up the input and began to generate a different chart.
These successfully generated charts were further classified into three categories based on the actions that the subjects performed to obtain them: (i) obtained on the first attempt, (ii) obtained after adjusting step or config settings, and (iii) obtained after rephrasing the input utterance.

Nearly half of the charts (23 out of 49, 47\%) were obtained on the first attempt. 
Besides, 13 cases involved step or config adjustments, and 13 cases involved input rephrasing. 
However, such adjustments did not necessarily imply that the system-generated results did not match their input. 
In fact, the input and the generated chart were consistent for all step or config adjustment cases and most rephrasing cases. 
Nonetheless, some subjects wanted to experiment with further adjustments after viewing the initial chart. 
We further counted the number of cases where the generated chart and user input matched among the rephrasing cases (10 out of 13). 
In the remaining three cases, the subjects attempted to rephrase their input once, twice, and thrice, respectively, until they obtained a result that matched their input.

Among the four failed inputs, three of them involved not supported data transformations or visual designs, such as dividing gross by budget or displaying two bar charts side by side in a single chart. 
The remaining input could not produce a valid chart as the provided encoding was self-contradictory (attempting to encode two data fields on the x-axis).

\subsubsection{Qualitative Feedback}

\textbf{The system's ability to respond to incomplete intent streamlines the thought process and enables users to explore data from the shallow to the deep. }
Most subjects involved some input with incomplete intent. 
Instead of referring to trends, distributions, or relationships, these inputs only indicate the data columns they are interested in, such as ``show some charts about major genre'' from S7 (left in \autoref{Fig:stage2}a1). 
S7 notes, ``When I first started looking at the data, I only had an initial interest in certain data columns (e.g. major genre).''
This allows them to give input once they have an initial idea and observe the system's response. 
S6 further mentioned, ``I only need to do one short step of thinking before viewing the results, while when using other tools, I often have to carefully define my intentions from vague to explicit.''
In addition, some subjects used the results from incomplete input to understand the data, draw connections, and develop further intents. 
For example, as is illustrated in \autoref{Fig:stage2}a2, after seeing the results of ``show horsepower'', S12 became interested in ``miles per gallon'' and entered ``describe horsepower and gallon''. 
Further, she wanted to focus on Japanese cars and typed ``show me the information about Japanese models related to gallon and horsepower'' and obtained the desired result. 
As such, the system's ability to answer abstract requests that don't articulate a complete intention shortens the thought process needed for every single round of interaction, enabling users to explore the data from the shallow to the deep. 

\textbf{ChartGPT supports a semantic understanding of the visual intent, allowing users to express themselves flexibly and naturally. }
Some subjects involved inputs that can not match the corresponding data columns directly. 
For example, when S6 entered ``which type of movies earn most'', the system could understand the keywords `type' and `earn' and infer the Major Genre and Worldwide Gross columns (middle in \autoref{Fig:stage2}a1).
Moreover, this semantic inference is not restricted to direct word-to-word mapping but is a general understanding. 
For example, on S8's input of ``number of movies over time,'' the system could determine that the `Release Year' column may be a more appropriate choice than `Running Time'. 
In this regard, about half of our subjects commented that the system is ``smart'' as it has some semantic inference ability and good support for natural language flexibility. 
Specifically, S2 praised its ``flexible semantic associations'' which alleviates his burden of perfecting their language to be more precise for the system. 
In general, our system's semantic understanding of utterances facilitates a more user-friendly experience as it reduces the need to be exact in users' phrasing. 

\textbf{The interaction to modify the results of intermediate steps can shorten the distance between the system-generated and user-desired results.}
Despite the majority of subjects recognizing ChartGPT's ability to understand semantic natural language and produce accurate results, due to user preferences and the ambiguity of the user's natural language, the generated results sometimes adhered to their expressions, yet did not yield their desired outcome in some parts.
For instance, S3 initially entered ``show the relationship between worldwide gross and rotten tomatoes rating'' and obtained a scatter plot between the two mentioned columns. 
However, she thought that this chart had too many points and wanted to focus on comedy movies, so she added the condition ``Major Genre = `Comedy''' to the filter step and regenerated the results (middle in \autoref{Fig:stage2}a3). 
S3 commented, ``I can regenerate results from the middle without reformulating my original input when I have a clear intent to target a particular step. ''
Overall, 10 of our 12 subjects have employed modifications of the steps or configurations according to their preferences. 
S2 further pointed out that after seeing the initial results generated by the system, it is simple to determine if its details match his preferences, resulting in a ``clear direction for modification''. 
\section{Discussion}

This section includes the implications, lessons learned, limitations, and future work of our system. 

\subsection{Implications}

In terms of technique, our framework employs LLMs for generating charts from abstract utterances using a ``decomposition and fine-tune'' approach that involves a limited-size dataset. 
We demonstrate its effectiveness through both quantitative evaluations and a user study. 
In terms of evaluation, we contribute a dataset of abstract utterances and corresponding charts generated using LLMs. 
This dataset can serve as a benchmark for future research and training data for machine learning studies. 
Additionally, our method of constructing the dataset from LLMs and using it to fine-tune LLMs is significant.
In terms of applicability, our framework's applicability extends beyond NL2VIS generation, as it can be used to solve complex downstream tasks that LLMs cannot directly handle. 
For instance, long story writing can also be decomposed into several sub-modules, from planning the characters and outline to drafting and editing the story continuation~\cite{yang2022re3}.
The feedback from these experiments provides valuable insights into the potential applications of LLMs in generating visualizations, inspiring further research in this field.

\subsection{Lessons Learned}
Modification is important to suit different preferences.
Users have varying preferences for chart design choices and may not always follow a consistent design rule. 
During our data collection, most of the data we collated tended to follow common design principles, such as using scatter plots for two quantitative data columns and line charts for displaying trends over time. 
However, our user study revealed that users' preferences were not always consistent. 
For instance, when aggregation was not specified explicitly in the utterances, some subjects preferred to average data while others preferred to look at the maximum value. 
Additionally, during the free exploration task, some subjects switched from a scatter plot displaying two quantitative data columns to a line graph or from a line graph showing trends over time to a bar graph. 
This underscores the importance of providing users with interactions to modify or fine-tune the results in the authoring tool to facilitate human-in-the-loop, as the generated results are not guaranteed to always match everyone's preferences.

\subsection{Limitations and Future Work}

\textbf{Support for a larger scope.}
ChartGPT only supports some key chart components and design choices for chart generation, with an aim to demonstrate the usefulness of our framework. 
Future work could involve support for a larger scope. 
First, additional transformations and visualization parameters could be considered.
Currently, we have not considered operations that reshape data tables, such as pivot and mutate~\cite{xiong2022revealing, kai2023somnus}.
We could add a transformation step before selecting the columns to support the transformations.
Parameters, such as mark types and visual channels, can be extended by enlarging our dataset.
Second, supporting follow-up utterances to modify the generated charts~\cite{mitra2022facilitating} is also an intuitive manner for human-LLM interaction.
To achieve this, we can train an LLM using a dataset with existing specifications and modification commands as input and an updated specification as output.
It requires the construction of the dataset, which can also be attained with the help of ChatGPT. 
Third, as an LLM for a specific domain, it is required to recognize out-of-domain queries and raise warnings.
To do so, we can add an additional boolean token representing whether the utterance is related to the input data and visual analysis.
Negative examples can be generated and mixed up with our proposed dataset.

\textbf{Scalability for large input tables.}
The model input includes table headers, column types, two data rows, and utterances. 
Therefore, the number of columns in the table would affect the prompt length. 
Based on our dataset, we trained the model with a maximum prompt length of 580 input tokens. 
To accommodate large tables exceeding this size, there are two potential improvements:
(a) Reconfigure the model input to reduce the token count.
For example, enabling LLMs to selectively choose columns, followed by the system providing additional values and information, can help reduce the prompt length. 
Such prompt improvement also holds the potential to provide deeper data insights, such as data distribution, within the constraints of a limited prompt length. 
(b) Expand our training dataset and allocate additional computational resources to accommodate longer prompts.
In addition, as a restricted prompt length would result in a less comprehensive inclusion of table information, further comparison with rule-based methods on large tables remains future work.

\textbf{Comparison with the generic LLM-based methods. }
Researchers have explored using generic LLM-based methods for chart generation.
For example, LIDA creates charts by generating and executing Python code. 
LIDA allows flexibility in selecting visualization libraries, such as Seaborn~\cite{Waskom2021seaborn} and Altair~\cite{VanderPlas2018altair}. 
Without a predefined design space, LIDA can accommodate diverse choices beyond our scope, posing challenges for applying our evaluation metrics. 
This limitation prompts the need for future research to compare such approaches comprehensively.

Despite the limitation, we tested LIDA on our test set. 
We observed 12 and 67 failed cases (raised errors while generating charts) for Seaborn and Altair, respectively (compared to 7 for ChartGPT). 
Many failures stem from calling nonexistent functions, revealing the inherent hallucination issues in generic LLMs.
In addition, LIDA's performance with Seaborn proves more stable than with Altair, possibly due to Seaborn's prevalence in the GPT corpus. 
This underscores generic LLMs' dependence on extensive existing corpora, exposing limitations in handling new schemas. 
Fine-tuning an LLM with specific data may compromise generalizability but can be valuable in scenarios requiring stability or new schemas.

\textbf{Inspiration v.s. accuracy. }
ChartGPT aims to accurately capture the intent from the user's abstract utterance and make reasonable inferences. 
Therefore, our system tends to prioritize the accuracy of the results by presenting the most relevant information first and providing optional charts later. 
Our dataset also reflects this tendency. 
When an utterance involves only a certain data column and lacks other intent, the ground truth is often to show the distribution of the column, which is the most closely related to the utterance's information. 

Despite this emphasis on accuracy, user feedback has indicated that it is not always the primary concern. 
For example, during the comparing and ranking task, for the utterance ``show something about origin,'' some of our subjects preferred the chart showing the origin and other data fields. 
Similarly, during the free exploring stage, three subjects suggested that they would like to see content that could inspire them beyond the scope of their utterances. 
Two of them emphasized that this requirement becomes more noticeable when seeking inspiration during data exploration.
This feedback shows a tendency for desiring charts that cover a broader range of data columns while exploring data~\cite {wongsuphasawat2015voyager1, wongsuphasawat2017voyager2}.
In response to this feedback, we plan to propose an option for users to specify their desired level of inspiration (e.g., ``high inspiration'' versus ``accuracy only'') in their query in the future. 
This allows the system to match users' needs better and enhance their experience.

\textbf{Flexibility v.s. certainty. }
Our system accommodated a wide range of user intentions, but limitations arose when users expressed intentions beyond the system's current capabilities. 
During our study, we observed two subjects attempting to explore data using intentions not supported by the system. 
One of the subjects expressed an intention that could not be drawn as a chart, while the other wanted to do a data transformation in which two columns in a table were computed, e.g., gross divided by budget. 
In such cases, our system still produced results, which unfortunately did not align with their intentions. However, it took the subjects quite some time to evaluate and finally realize that the system did not support their intentions after adjusting their inputs several times. 
While our design space could be expanded to accommodate more needs, the flexibility of natural language and the definite design space of the system mean that the system's capability is limited to support the full range of natural language expressions, leading to confusion for users about which inputs will lead to successful chart results.
Future work could explore enhancing the system's recognition of inputs beyond its supported range. 
For example, instead of implementing all possible user intentions beyond the scope, one potential avenue is to integrate a preliminary step that identifies inputs exceeding the system's limits and issues a warning.

\section{Conclusion}

This paper introduces ChartGPT, leveraging LLMs to generate charts from abstract utterances.
We formulate the chart-generation problem as a sequential reasoning task and 
construct an abstract utterance dataset to fine-tune a language model for solving each task. 
Furthermore, we design an interactive interface for ChartGPT to enable users to examine and modify intermediate outputs. 
The effectiveness of the proposed system is evaluated through comparative and usability studies.

\section*{Acknowledgments}
This work was supported by NSFC (U22A2032) and Key ``Pioneer'' R\&D Projects of Zhejiang Province (2023C01120).

\bibliography{ref}
\bibliographystyle{IEEEtran}

\begin{IEEEbiography}[{\includegraphics[width=1in,height=1.25in,clip,keepaspectratio]{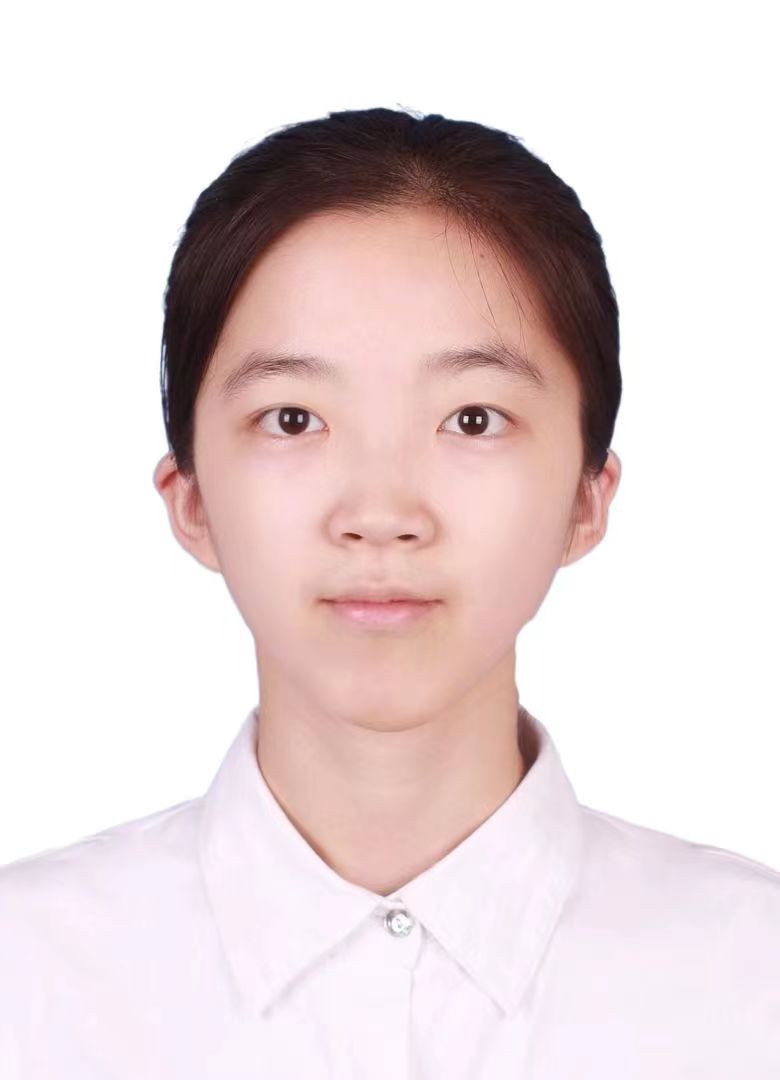}}]{Yuan Tian}
received her B.S. degree in computer science from Zhejiang University in 2022. She is currently a Ph.D. student in the State Key Lab of CAD\&CG, Zhejiang University. Her research interests include machine learning for visualization and visual analytics. 
\end{IEEEbiography}

\begin{IEEEbiography}[{\includegraphics[width=1in,height=1.25in,clip,keepaspectratio]{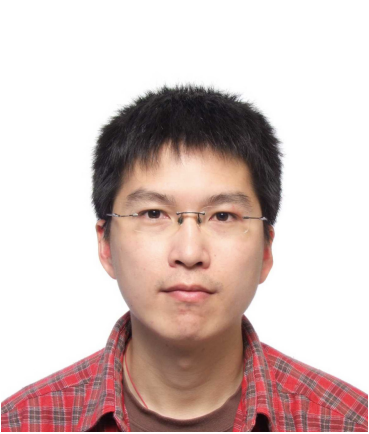}}]{Weiwei Cui} received the BS degree in computer science and technology from Tsinghua University, China, and the PhD degree in computer science and engineering from the Hong Kong University of Science and Technology, Hong Kong. He is a principal researcher at Microsoft. His primary research interest is visualization, with the focuses on democratizing visualization and AI-assisted design. For more information, please visit \url{https://www.microsoft.com/en-us/research/people/weiweicu/}.
\end{IEEEbiography}

\begin{IEEEbiography}
[{\includegraphics[width=1in,height=1.25in,clip,keepaspectratio]{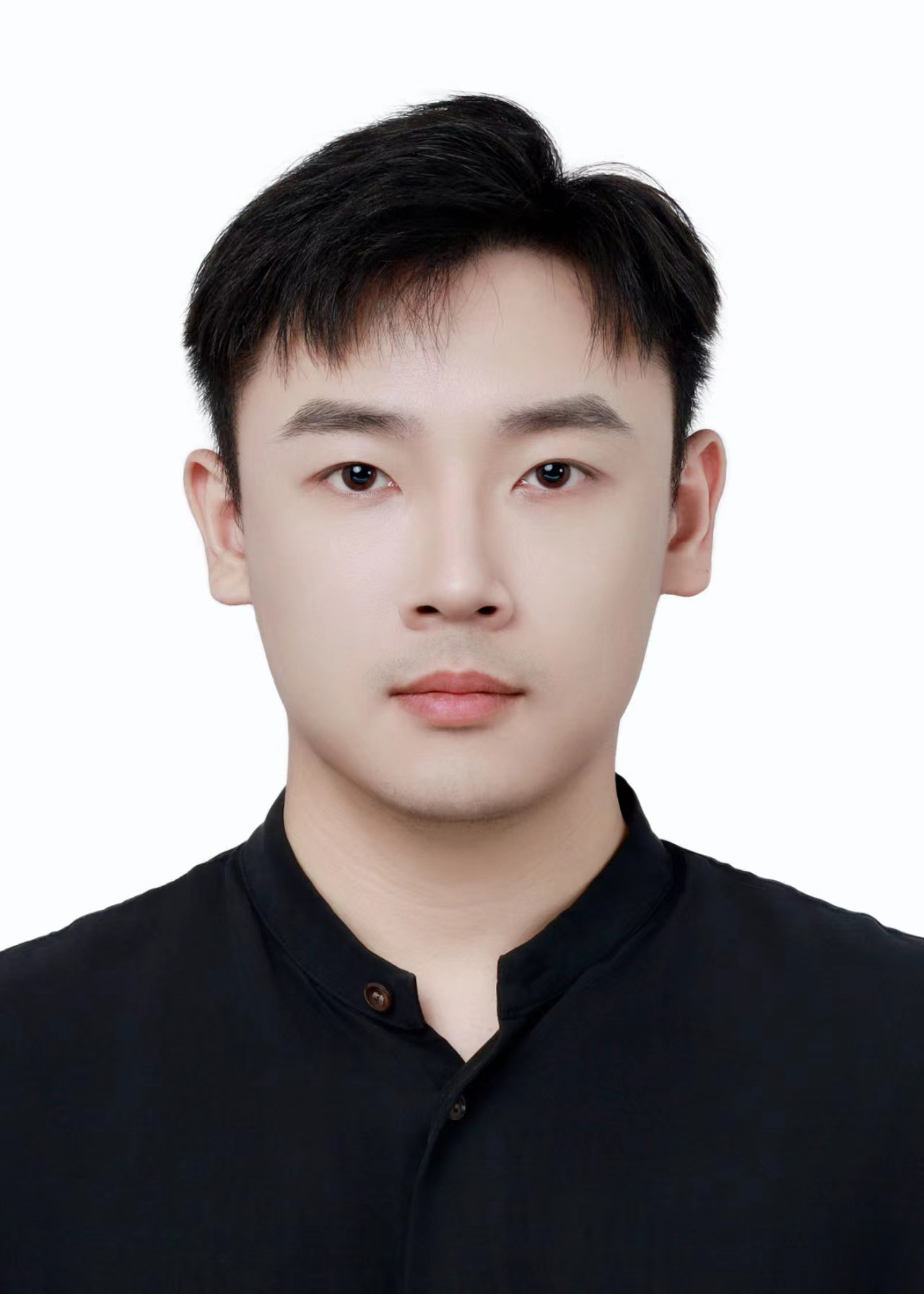}}]{Dazhen Deng} 
is currently a tenure-track assistant professor at the School of Software Technology, Zhejiang University. He received Ph.D. in Computer Science from Zhejiang University in 2023. His research interests mainly lie in machine learning for visual analytics. For more information, please visit https://dengdazhen.github.io/.
\end{IEEEbiography}

\begin{IEEEbiography}
[{\includegraphics[width=1in,height=1.25in,clip,keepaspectratio]{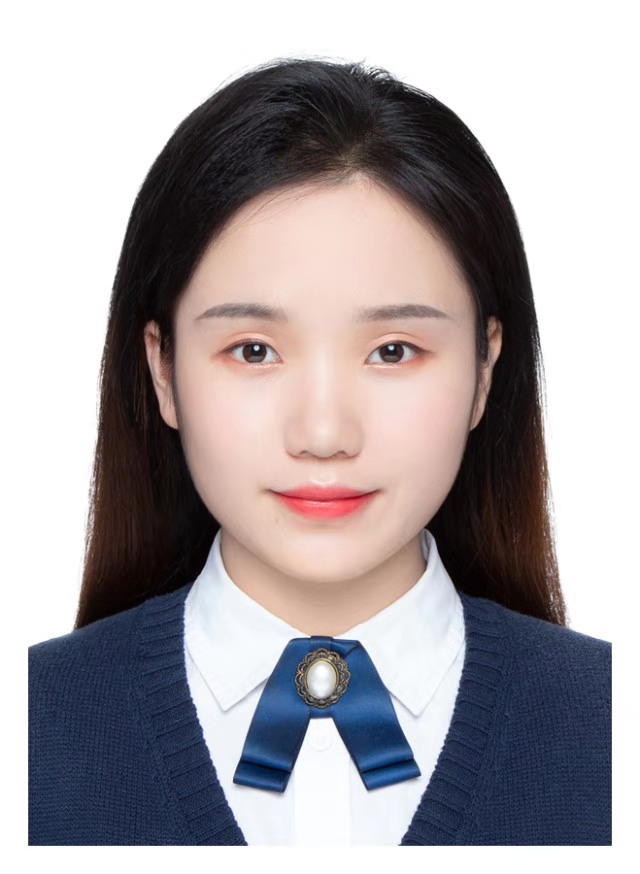}}]{Xinjing Yi} received her B.S degree in computer science from Wuhan University in 2022. She is currently a graduate student in Software Engineering, Zhejiang University. Her research interests mainly include visualization and visual analytics.
\end{IEEEbiography}

\begin{IEEEbiography}[{\includegraphics[width=1in,height=1.25in,clip,keepaspectratio]{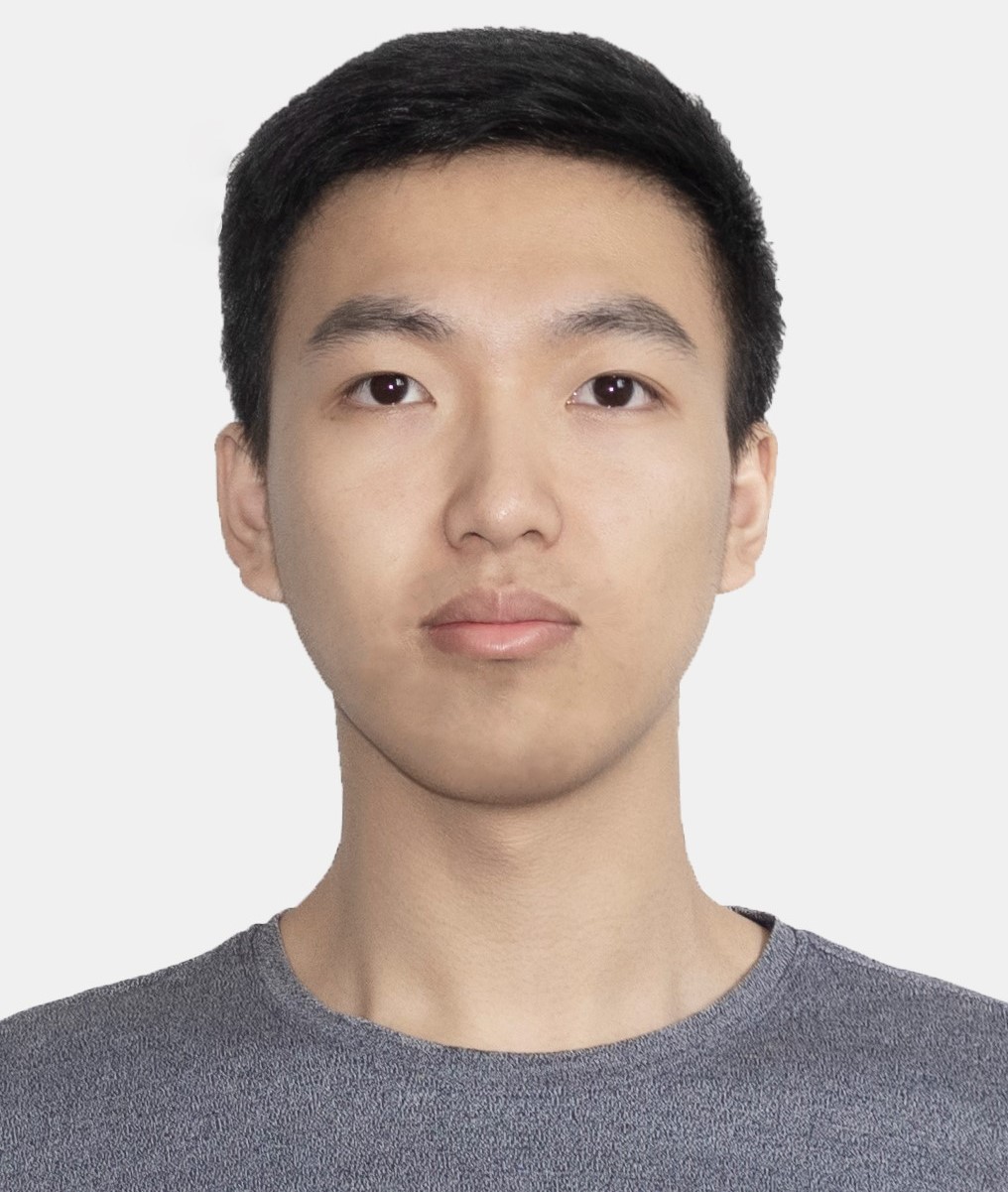}}]{Yurun Yang} received his B.S. degree in Software Engineering from University of Electronic Science and Technology of China in 2022. He is currently a graduate student in School of Software Technology, Zhejiang University. His research interests mainly include visualization and visual analytics.
\end{IEEEbiography}

\begin{IEEEbiography}[{\includegraphics[width=1in,height=1.25in,clip,keepaspectratio]{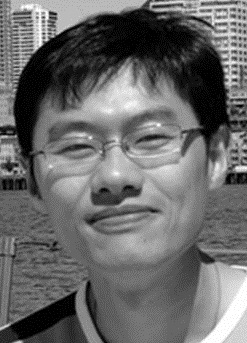}}]{Haidong Zhang} received the PhD degree in Computer Science from Peking University, China. He is a Principal Architect at Microsoft. His research interests include visualization and human-computer interaction.
\end{IEEEbiography}

\begin{IEEEbiography}[{\includegraphics[width=1in,height=1.25in,clip,keepaspectratio]{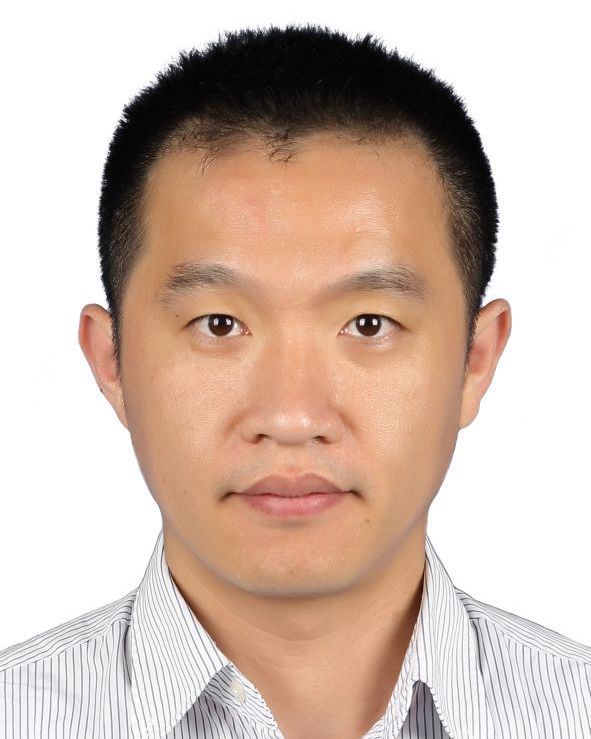}}]{Yingcai Wu} is a Professor at the State Key Lab of CAD\&CG, Zhejiang University. His main research interests are in information visualization and visual analytics, with focuses on urban computing, sports science, immersive visualization, and narrative visualization. He received his Ph.D. degree in Computer Science from the Hong Kong University of Science and Technology. Prior to his current position, Dr. Wu was a postdoctoral researcher in the University of California, Davis from 2010 to 2012, a researcher in Microsoft Research Asia from 2012 to 2015, and a ZJU100 Young Professor at Zhejiang University from 2015 to 2020. For more information, please visit \url{http://www.ycwu.org}.
\end{IEEEbiography}

\vfill

\end{document}